%% file: main.tex
\documentclass[
aip,
graphicx,
amsmath,
amssymb,
reprint,
floatfix]{revtex4-1}

\usepackage{graphicx}
\usepackage{helvet}
\usepackage{xurl}
\usepackage{hyperref}
\usepackage{amsmath} 
\usepackage{amssymb} 
\usepackage{orcidlink}
\usepackage{comment}
\usepackage{cleveref}
\usepackage{float}
\usepackage{svg}
\usepackage{dcolumn}
\usepackage{bm}
\usepackage{tabularx}
\usepackage{placeins}

\usepackage[utf8]{inputenc}
\usepackage[T1]{fontenc}

\newcommand{\myCite}[1]{\citeauthor{#1} \cite{#1}}

\draft 

\begin{document}


\title{Improving conditional generative adversarial networks for inverse design of plasmonic structures}
\author{Petter Persson}
\author{Nils Henriksson}
\author{Nicolò Maccaferri}
\email{nicolo.maccaferri@umu.se}
\affiliation{Department of Physics, Umeå University, 901 87 Umeå, Sweden}


\date{\today}

\begin{abstract}
\input{chapters/abstract}
\end{abstract}


\maketitle 

\section{Introduction}
\input{chapters/introduction}

\section{Methods}
\input{chapters/methods}

\section{Results and Discussion}
\input{chapters/results_discussion_conclusion}

\section{Acknowledgments}
\input{chapters/acknowledgments}

%
%

%


\clearpage
\section{References}
\bibliography{chapters/references.bib}

\newpage
\section{Supplementary Material}
\input{chapters/supplementary}

\end{document}

%% file: chapters/abstract.tex
Deep learning has emerged as a key tool for designing nanophotonic structures that manipulate light at sub-wavelength scales. We investigate how to inversely design plasmonic nanostructures using conditional generative adversarial networks. Although a conventional approach of measuring the optical properties of a given nanostructure is conceptually straightforward, inverse design remains difficult because the existence and uniqueness of an acceptable design cannot be guaranteed. Furthermore, the dimensionality of the design space is often large, and simulation-based methods become quickly intractable. Deep learning methods are well-suited to tackle this problem because they can handle effectively high-dimensional input data. We train a conditional generative adversarial network model and use it for inverse design of plasmonic nanostructures based on their extinction cross section spectra. Our main result shows that adding label projection and a novel embedding network to the conditional generative adversarial network model, improves performance in terms of error estimates and convergence speed for the training algorithm. The mean absolute error is reduced by an order of magnitude in the best case, and the training algorithm converges more than three times faster on average. This is shown for two network architectures, a simpler one using a fully connected neural network architecture, and a more complex one using convolutional layers. We pre-train a convolutional neural network and use it as surrogate model to evaluate the performance of our inverse design model. The surrogate model evaluates the extinction cross sections of the design predictions, and we show that our modifications lead to equally good or better predictions of the original design compared to a baseline model. This provides an important step towards more efficient and precise inverse design methods for optical elements.

%% file: chapters/introduction.tex
\noindent
Many novel ways to manipulate the interaction between light and matter have been discovered during the past decades, which has been of great importance in nanophotonics research and applications \cite{gonzalez2024light,maccaferri2021recent,koya2023advances}. New materials and structures, such as metamaterials and plasmonic antennas, can be designed to achieve desired optical properties and are used in a wide range of applications, including plasmonic curcuits \cite{timothy2017plasmonic,tuniz2020modular}, super resolution imaging \cite{willets2017super} and sensing \cite{mesch2016nonlinear,mayer2011localized,langer2020present}. As the fabrication possibilities grow, so does the number of available design parameters \cite{brown2019machine}. The complexity of the design problem increases with the size of the design parameter space, leading to several challenges when developing methods to predict optimal designs with respect to the desired optical properties. These challenges include limitations in existing physics-based approaches to model light-matter interactions of particles with complex geometry, leading to a dependence on numerical simulations of electromagnetic models instead. Numerical simulations are limited by the available computational resources, which puts a constraint on both memory usage and time complexity of the design algorithm, making it a challenge to employ optimization techniques that handle large amounts of design parameters \cite{molesky2018inverse, ma2021deep}. Therefore we need smarter, more efficient approaches for inverse design, and in this work we develop a machine learning framework for this purpose.

Various machine learning approaches have been tested for nanophotonic inverse design applications \cite{molesky2018inverse,brown2019machine,moon2023review,masson2023machine}. A particularly effective machine learning approach has been to use deep learning models based on neural networks to automate the design process. The main advantage of the deep learning approach is that it provides versatile methods that allows for learning arbitrary, non-linear continuous functions between finite-dimensional sets of inputs and outputs. This is possible since neural networks act as universal function approximators in the sense that, there exist a neural network to approximate any continuous function to a desired accuracy. Formally, any Borel measurable function can be approximated with a neural network given that there are enough hidden units available in the layers of the network \cite{hornik1989multilayer}. Furthermore, deep learning models excel at handling high-dimensional data, and there is a broad range of applications, including image analysis and non-linear regression tasks that are highly relevant for nanophotonics \cite{lecun2015deep}. Therefore, it is clear that this family of methods are well for the inverse design problem in nanophotonics \cite{so2020deep}.
\begin{figure*}[htb]
    \centering
    \begin{minipage}{0.48\textwidth}
        \centering
        \includegraphics[width=\linewidth]{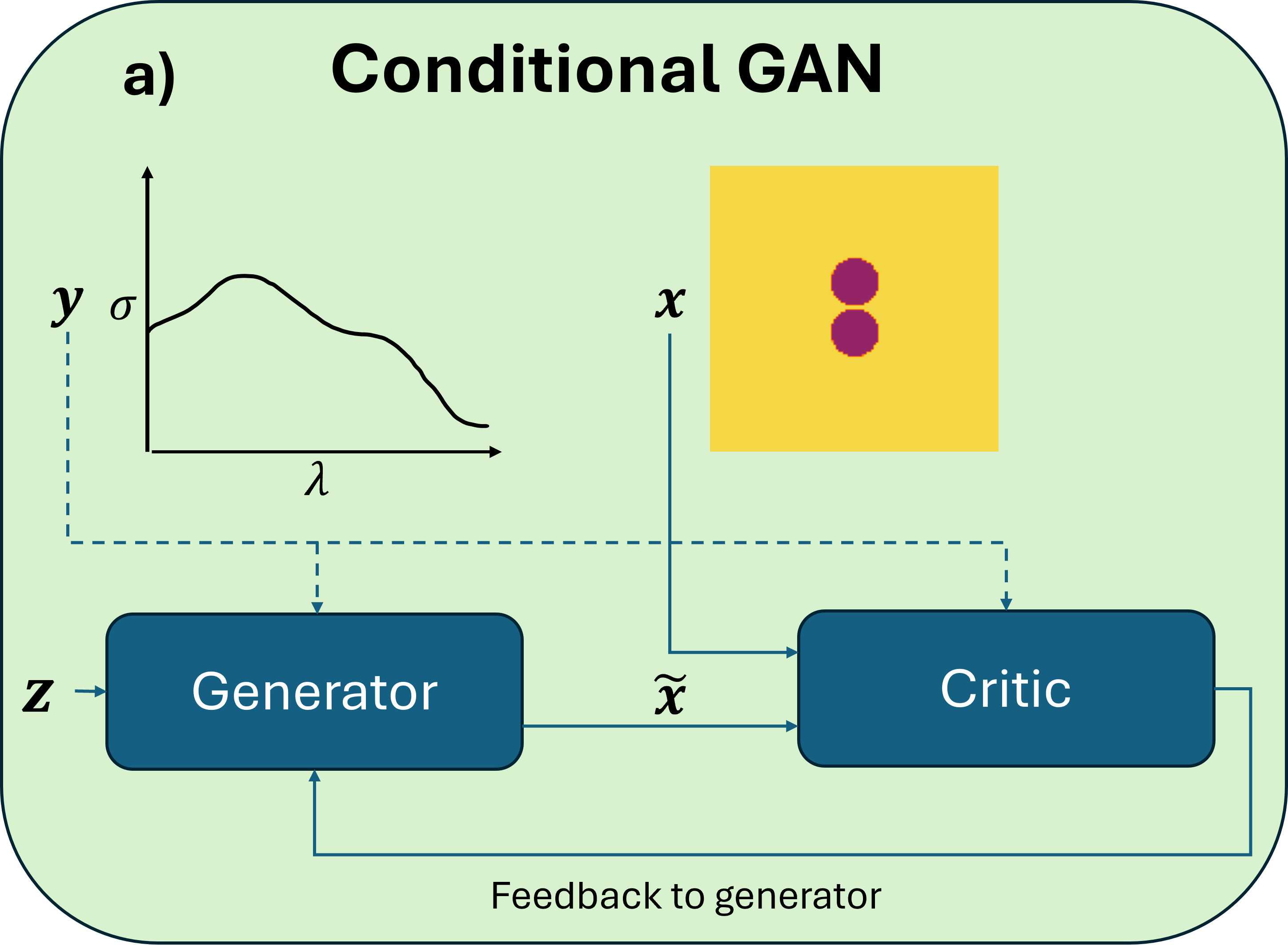}
    \end{minipage}
    \hfill
    \begin{minipage}{0.48\textwidth}
        \centering
        \includegraphics[width=\linewidth]{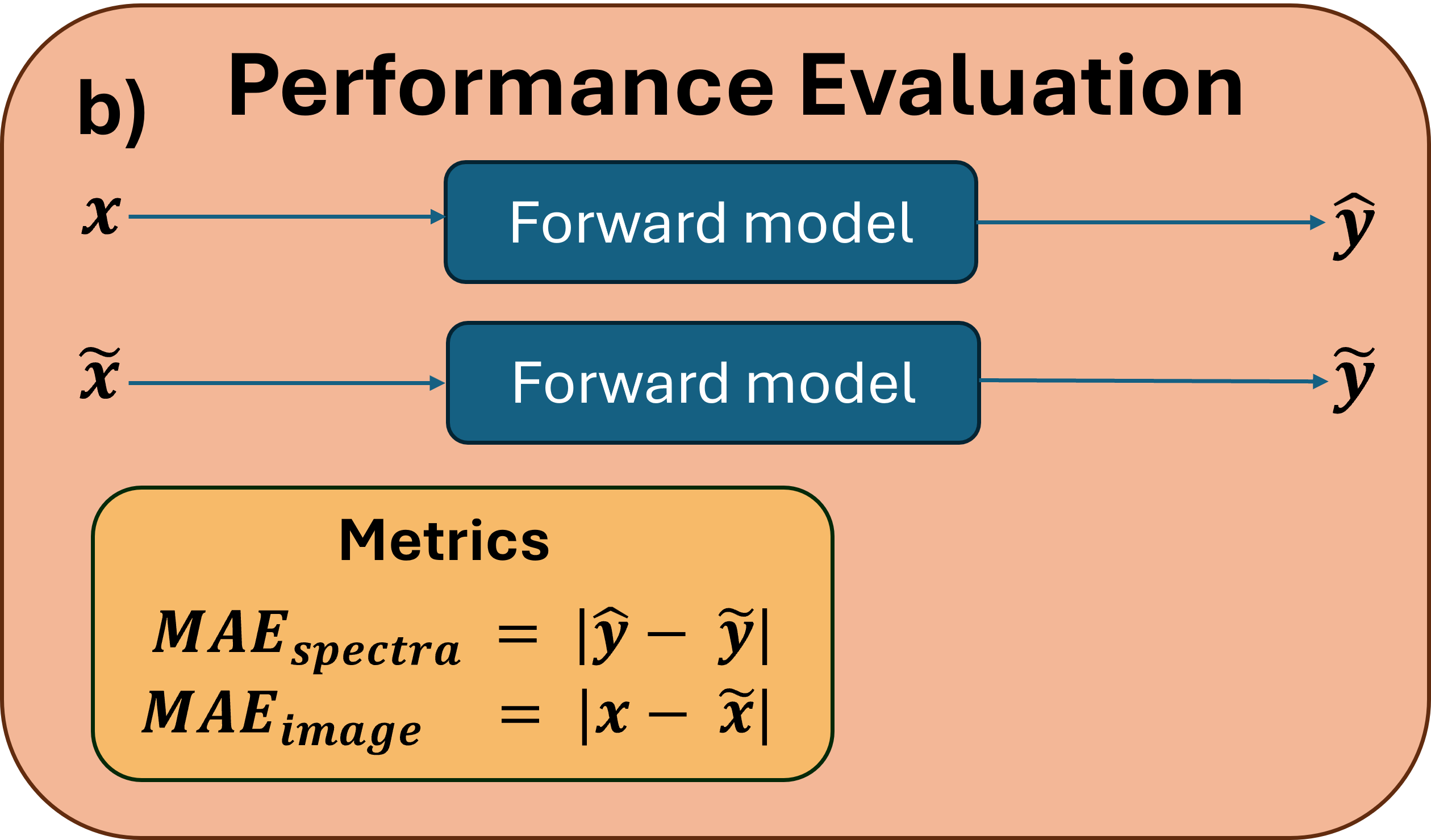}
    \end{minipage}

    \caption{This schematic presents an overview of the inverse design framework presented in this work. \textbf{a)} We train a cGAN-model consisting of a generator and a critic network, to predict optimal designs of dimer structures. Based on the desired cross-section spectra $\mathbf{y}$, and a stochastic vector $\mathbf{z}$, the generator learns to predict the design $\mathbf{\tilde{x}}$. The critic network provides feedback to the generator by estimating the statistical distance between the original design $\mathbf{x}$ and $\mathbf{\tilde{x}}$ produced by the generator. The model is empirically optimized to minimize the statistical distance between samples, using simulation data. Training is stopped when the distribution of $\mathbf{\tilde{x}}$ has converged to the distribution of $\mathbf{x}$. \textbf{b)} We pre-train a CNN based forward model to predict the spectra $\mathbf{tilde{y}}$ associated with a generated design and use it for performance evaluation. The forward model allows for estimating the mean absolute error (MAE) between the spectra of the generated design and the original spectra, providing a quantitative measure of their similarity. In addition, we evaluate the image based MAE of the predicted design as well.}
    \label{fig:1}
\end{figure*}

Using a data-driven approach, the inverse design models are optimized using synthetic data generated from electromagnetic simulations of the problem at hand. Promising implementations include convolutional neural networks (CNN) as forward models predicting the optical response of a plasmonic particle \cite{sajedian2019finding,noureen2021deep}, a tandem deep neural network (DNN) combining the forward model with an inverse network for predicting designs \cite{liu2018training}, conditional autoencoder (cAE) models \cite{ma2019probabilistic, ma2020data}, and conditional generative adversarial networks (cGAN) \cite{so2019designing,an2021multifunctional,liu2018generative,jiang2019free}. The forward model approach is usually combined with another optimization technique, such as evolutionary optimization algorithms. In this case, the forward model acts as a surrogate model for ultra-fast evaluation \cite{chugh2019surrogate, feichtner2017plasmonic}. In contrast, tandem DNN, cAE, and cGAN models are examples of direct inverse design modeling using neural networks. With this approach, the main challenge to address is the non-uniqueness of solutions called the one-to-many problem. Since many inverse design problems lack a unique solution, there may exist several solutions or even no solution to a given spectral input \cite{wiecha2021deep}.

While many deep learning techniques have been applied to improve design rules in nanophotonics, the focus has been mostly on finding a model that works for the particular application of interest, and not on the optimization. Hence, this work focuses on how to improve existing models typically used by the nanophotonics community. In particular, we study the conditional GAN model to design plasmonic nanostructures of anisotropic shape, including plasmonic dimers. For such structures, the optical response depends on the polarization of the incoming light \cite{Verre2016}, making them relevant for studying plasmonic effects in the near-field. We make two significant optimizations to the model that increase its accuracy and its convergence rate. The improvements are shown for two different network architectures. One simple feed-forward neural network architecture consisting of fully connected layers, that we call the fully connected GAN model (FCGAN) \cite{sazli2006brief}. Second, we propose a more complex deep convolutional GAN model (DCGAN) introduced by \myCite{radford2015unsupervised} and used by \myCite{so2019designing}, for the inverse design of silver antennas. We also apply Wasserstein GAN (WGAN) loss and training algorithm to stabilize the optimization that is inherently unstable, often leading to convergence issues \cite{gulrajani2017improved}. Figure \ref{fig:1} presents an overview of our inverse design framework based on the cGAN-model. The goal of cGAN-model is to learn a generator function that maps from a stochastic input space to the design space conditioned on the desired optical spectra. Given the spectra $\bm{y}$ and the stochastic vector $\bm{z}$ it outputs a prediction of the optimal design $\bm{\tilde{x}}$. During training, a critic network provides feedback to the generator by estimating the statistical distance between the distribution of original and predicted design images, and the model is optimized to minimize this quantity. 

We use a surrogate-model-based metric to evaluate the performance of the cGAN-model. The surrogate model uses a pre-trained forward model based on a CNN-architecture, which predicts the spectra associated with a given design image. This allows for estimating the mean absolute error (MAE) in the spectra by using the forward model to predict the spectra of both the original and generated designs. In addition to the surrogate model, we also use an image-based performance metric. The MAE is evaluated for the actual design images, i.e. between $\bm{x}$ and $\bm{\tilde{x}}$, to evaluate the pixel to pixel loss in the predicted designs.

%% file: chapters/methods.tex
\noindent Conditional GAN-models have commonly been used for inverse design problems in nanophontonics \cite{so2019designing, an2021multifunctional}. Specifically, the cGAN model is used to learn the relationship between the geometry of a structure and its optical properties. We use the scattering and absorption cross section spectras as the conditional input. The model consists of two networks, a generator $G$, and a discriminator $D$, that are trained together to gradually improve their performance \cite{goodfellow2014generative}. In general, the inverse design problem does not have unique solution, but cGAN-models handle this well because the generator model $G$ takes a stochastic input as well, making it possible to learn several solutions to a single optical spectra. For this reason, the cGAN-model have become a popular choice when inversely designing nanostructures to satisfy the users desired optical properties.

In our study, we utilize the cGAN-model to generate nanostructure geometries $\bm{x}$, conditioned on the structures scattering and absorption cross-section spectra $\bm{y}$. The goal is to learn a generator model $\bm{x} = G(\bm{z}, \bm{y}) \sim P_g$ to mimic the training data distribution $P_r$, from taking a stochastic vector $\bm{z}$, conditioned on the label vector $\bm{y}$, as model input. Furthermore, we use the Wasserstein loss function to optimize our model, as suggested by \myCite{gulrajani2017improved}. In this case the discriminator is called the \textit{critic} since it measures the statistical distance between true $P_r$ and generated distributions $P_g$.

\begin{figure*}[htb] 
    \centering 
    \includegraphics[width=0.8\textwidth]{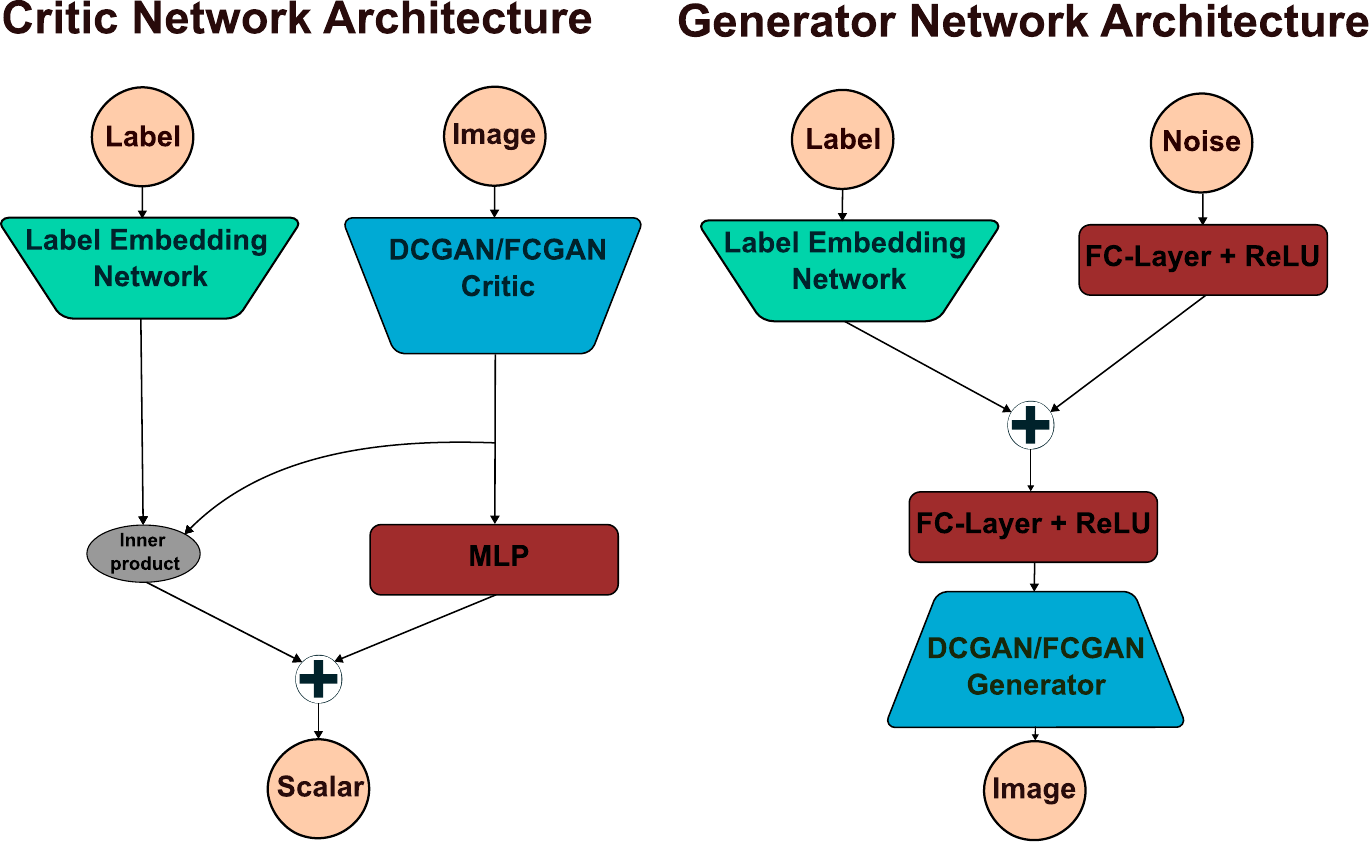} 
    \caption{The general architecture of the conidtional GAN model consists of critic network and generator network. The critic network is used by the Wasserstein GAN training algorithm in the optimization process. During the optimization process, the generator network is trained to generate images that replicate the probability distribution of the training data. Both networks takes conditional label data as input. The label data is the absorption and scattering cross section spectral data of the corresponding training images used to train the model. Proposed in this work is to introduce the conditional label data using label projection in the critic network. The label vector is projected onto the feature vector of the image processing network (blue) with an inner product and this is called label projection. Furthermore we also propose the use of a label embedding network to encode the label data. The label embedding network uses several layers of the one-dimensional convolutions.} \label{fig:network_architecture} 
\end{figure*}

Our model architecture is illustrated in \cref{fig:network_architecture}. We use an improved method to input the conditional data into the critic model, based on the label projection method proposed by \cite{miyato2018cgans}. The method introduces the conditional data into the model with an inner product, instead of concatenation or vector addition. This projection-based method is more consistent with the underlying probabilistic model of the standard adversarial discriminator loss \cite{goodfellow2014generative} and \myCite{miyato2018cgans} show that it improves the generation of class conditional images. Thus, we assume that similar performance gains can be achieved when using the Wasserstein loss function \cite{gulrajani2017improved}. In addition to label projection, we also add capacity in the critic using a label embedding network to further improve its ability to learn features from the conditional input. The label embedding network consists of multiple one-dimensional convolutional layers to process the input data, and the full cGAN-model is mathematically formulated as 
\begin{equation}
\begin{aligned}
    & D:\mathbb{R}^s \times \mathbb{R}^{c \times h \times \times w} \rightarrow \mathbb{R} \\
    & G:\mathbb{R}^s \times \mathbb{R}^{z} \rightarrow \mathbb{R}^{c \times h \times \times w}
\end{aligned}
    \label{eq:GAN}
\end{equation}
where
\begin{equation}
    D(\bm{x,y}) = \mathrm{MLP}(\bm{f}(\bm{X})) + \bm{f}(\bm{x})^\top \bm{e}(\bm{y})
    \label{eq:disc}
\end{equation}
is the critic model, and
\begin{equation}
    G(\bm{z},\bm{y}) = \bm{g}(\bm{z} + \bm{e}(\bm{y}))
    \label{eq:gen}
\end{equation}
is the generator model. Both are parametrized with neural networks and optimized with respect to \cref{eq:wgan}. The critic model takes an image $\bm{x} \in \mathbb{R}^{c \times h \times w}$ and spectral data $\bm{y} \in \mathbb{R}^{s}$ as input, and outputs a scalar. It contains the image downsampling network $\bm{f}: \mathbb{R}^{c \times h \times w} \rightarrow \mathbb{R}^{d_e}$, and the embedding network $ \bm{e}: \mathbb{R}^s \rightarrow \mathbb{R}^{d_e}$, which downsample the spectral data into a lower dimensional latent space. The dimension of the latent space is chosen as $d_e = 50$. \Cref{tab:emb} describes the architecture of the embedding network. Batch normalisation is used after each convolutional layer, together with an activation function. For consistency with the downsampling networks of the two architectures, we use ReLU as activation in the FCGAN-model, and LeakyReLU with a slope of 0.2 in the DCGAN-model. The second term in \cref{eq:disc} is the label projection term, where the conditional information is introduced with a euclidean inner product. A small multilayer perceptron (MLP) is also used to make the dimension compatible with the label projection term. The generator takes a Gaussian vector \textbf{z} $\sim \mathcal{N}(\mathbf{0}, I) \in \mathbb{R}^{z}$, where size of $\bm{z}$ is 100, and spectral data $\bm{y} \in \mathbb{R}^s$ as input and outputs an image $\bm{x} \in \mathbb{R}^{c \times h \times w}$. The generator upsampling network is denoted by $\bm{g} \in \mathbb{R}^{c \times h \times w}$. During the development of our model, we found that including the label embedding network in the generator improved the performance of our models. This is a logical consequence, as the additional model parameters allows the generator to learn a richer set of features from the input spectra, leading to more precise predictions of nanostructure geometries. Therefore, the embedding network was included in the generator architecture as well.
\begin{table}[htb]
\centering
\caption{The table summarize the label embedding network used in the GAN-models.}
\newcolumntype{C}{>{\centering\arraybackslash}X}
\begin{tabularx}{\linewidth}{|c|C|c|c|}
\hline
Stage & Operation                                                                     & Channels    & \begin{tabular}[c]{@{}c@{}}Kernel\\ /Stride\\ /Padding\end{tabular} \\ \hline
1     & \begin{tabular}[c]{@{}l@{}}Conv1D\\ + BatchNorm1D\\ + Activation\end{tabular} & 32          & 5/2/0                 \\ \hline
2     & \begin{tabular}[c]{@{}l@{}}Conv1D\\ + BatchNorm1D\\ + Activation\end{tabular} & 64          & 5/2/2                 \\ \hline
3     & \begin{tabular}[c]{@{}l@{}}Conv1D\\ + BatchNorm1D\\ + Activation\end{tabular} & 128         & 5/2/2                 \\ \hline
4     & \begin{tabular}[c]{@{}l@{}}Conv1D\\ + BatchNorm1D\\ + Activation\end{tabular} & 512         & 3/2/1                 \\ \hline
5     & Average Pool                                                                  & -           & -                     \\ \hline
      &                                                                               & Input dim & Output dim          \\ \hline
6     & FC-layer                                                                      & 512         & $d_e$                   \\ \hline
\end{tabularx}
\label{tab:emb}
\end{table}
\noindent For the critic and the generator, we study two different neural network architectures. First, we show that a simple non-convolutional model using fully connected layers exclusively, benefit from our proposed improvements. We call this the FCGAN-model, where both the critic and the generator have 3 hidden layers with 512 neurons each, and a ReLU activation function after each hidden layer. Second, we also study our model on the deep convolutional architecture proposed by \cite{radford2015unsupervised} and used in the inverse design study by \myCite{so2019designing}. The network architectures details can be found in the Supplementary Material \cref{sec:supp}.

We use the Wasserstein distance in the loss function with a gradient penalty term, as it has been shown to provide smoother gradients that stabilize the optimisation process \cite{gulrajani2017improved}. The loss is summarized as
\begin{align}
\begin{split}
    L = &\mathbb{E}_{\bm{\tilde{x}} \sim P_g} [D(\bm{\tilde{x}} | \bm{y})] - \mathbb{E}_{\bm{x} \sim P_r} [D(\bm{x} | \bm{y})] + \\ &\lambda \mathbb{E}_{\bm{\hat{x}} \sim P_{\bm{\hat{x}}}} \left( (\|\nabla_{\bm{\hat{x}}} D(\bm{\hat{x}} | \bm{y})\|_2 - 1)^2 \right) 
    \label{eq:wgan}
\end{split}
\end{align}
where the critic $D(x|y)$ and the generator $G(z|y)$ are parametrized with neural networks. \myCite{gulrajani2017improved} enforce a soft version gradient penalty term, where $\hat{x}$ is an interpolation between a real and generated sample. The hyperparameter $\lambda$ controls the size of the penalty term, and we use $\lambda = 10$ in our study, as proposed by \myCite{gulrajani2017improved}. Our model is optimized with the proposed training algorithm from \cite{gulrajani2017improved}.

%% file: chapters/results_discussion_conclusion.tex
\begin{figure*}[htb!]
    \centering
    \includegraphics[width=0.65\linewidth]{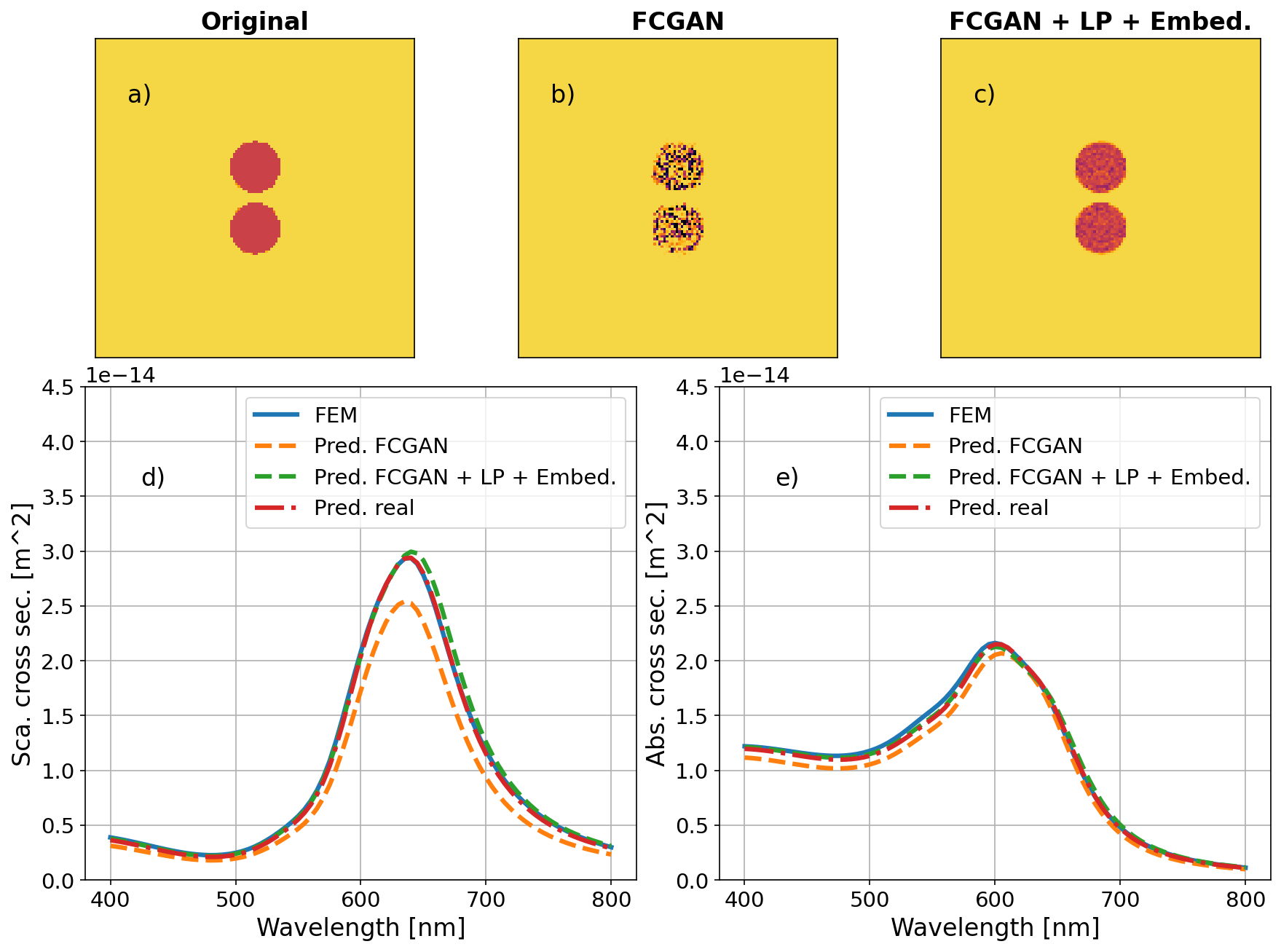}
    \caption{The results in this figure are obtained from the FCGAN-model with and without using label projection and an embedding network in the critic. Figure a) show the original image and figures b) and c) shows the images predicted by the models. Figures d) and e) show the scattering and absorption cross section spectras corresponding to the images. The blue curve is the original data from the finite element simulation of the gold nano structure in the original image. We use a pre-trained convolutional neural network regression model to predict the spectral data for both the original image (dashed red) and the images generated by the model (dashed green and orange). The quality is visibly worse for the image generated from FCGAN compared to when we add label projection and the embedding network. This is also causing a pertubation in the spectral data prediction.}
    \label{fig:2}
\end{figure*}
\noindent Our proposed cGAN models were trained on a dataset consisting of anisotropically shaped plasmonic nanostructures. The dataset includes both dimers and structures of elliptical shape. The structures are made of gold and placed on top of a glass substrate, which defines the domain of size 500x500 nm$^2$ in cross-section. When creating the dataset, the aim was to vary the available geometrical parameters, while simultaneously limiting any possible bias. Therefore, we varied the height, diameter and dimer gap using uniform distributions to randomly generate different dimers. The polarization of the light was always 45° to excite both the parallel and the perpendicular axis of the dimers \cite{Verre2016}. We used a steady state simulation of each generated structure using the finite element method (FEM) and calculated the absorption and scattering cross-sections. The COMSOL Multiphysics \textsuperscript{\tiny\textregistered} software and its wave optics module was used to perform the simulations \cite{comsol2025}. In the dataset, the dimer gap ranges from 5-40 nm, the long axis length scale from 40-100 nm, while the height scale is 14-146 nm. We employ the Brendel-Bormann model \cite{rakic1998optical} for the permittivity of gold to simulate the electromagnetic response of the structures. All structures are simulated for wavelengths in the range 400-800 nm, using 16 nm steps. In total, the dataset contained 2898 samples of different structures with the corresponding spectral data. 
\begin{figure*}[htb!]
    \centering
    \includegraphics[width=0.6\linewidth]{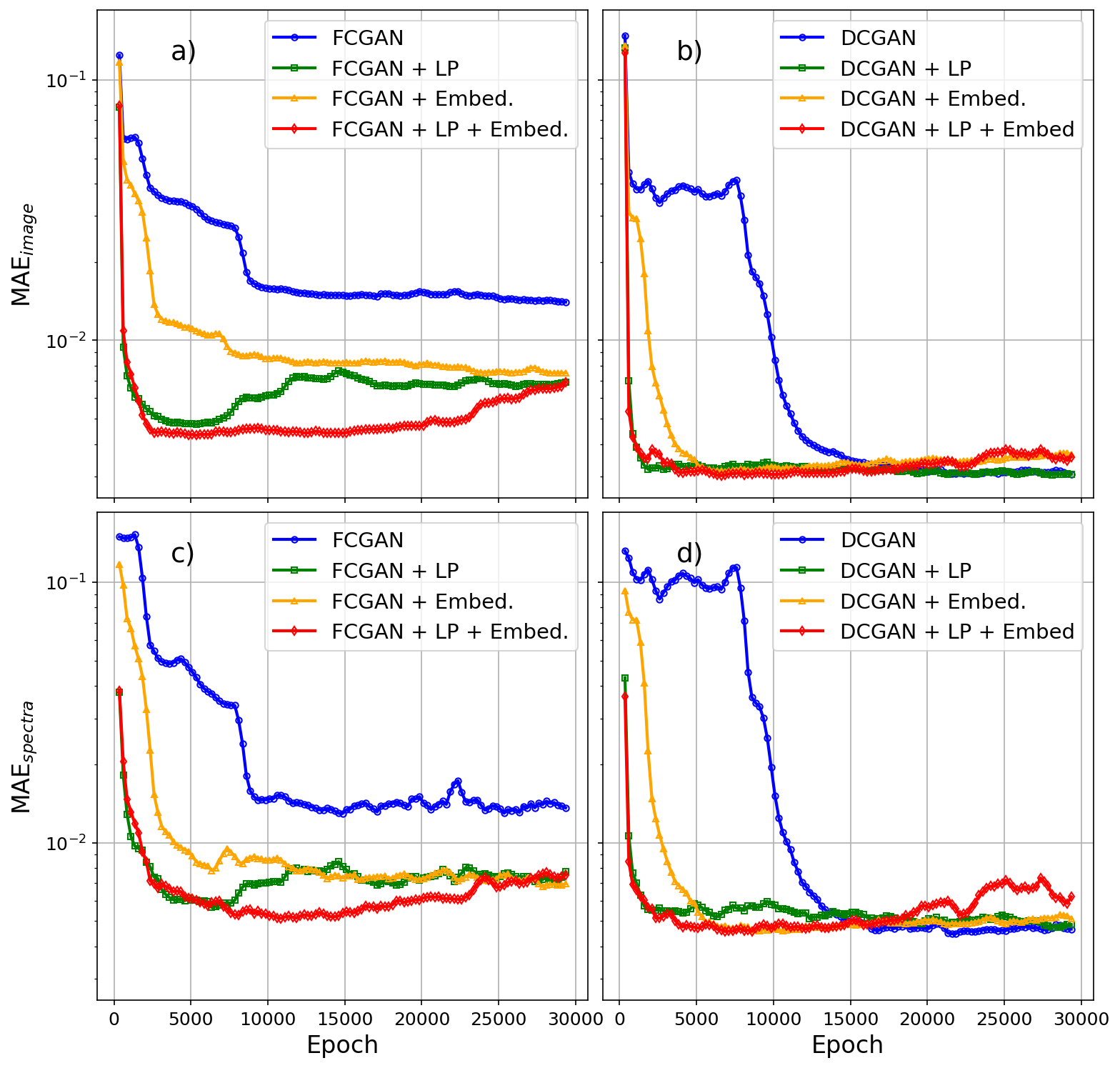}
    \caption{The models were trained on a dataset containing cylindrical dimer structures and they were evaluated using two different methods. The first method uses a pre-trained convolutional neural network regression model to predict the scattering cross section and absorption cross section spectra. This approach allows us to estimate the MAE between the spectra predicted from real and GAN-generated images. As a second evaluation method, we also estimate the mean absolute error in pixel values between real and generated images. Figures a) and c) shows the results from training the FCGAN architectures figures b) and d) shows the results from training the DCGAN-architecture. We calculate the evaluation metrics by generating n=30 sample images for each labeled original image, and take the average of those when computing the error estimate.}
    \label{fig:3}
\end{figure*}

We used the Wasserstein GAN training algorithm \cite{gulrajani2017improved} to train the models for 30000 epochs on the dimer dataset, and both the FCGAN and the DCGAN architectures are used in the critic and the generator networks. The effect of adding label projection and the embedding network is evaluated in a comparison study. Figure \ref{fig:3} shows the progression of the validation error during training, calculated on a separate validation set not used by the optimisation algorithm. From the validation set, we estimated the MAE in two different ways and use it as our metric for model performance. For the first method, we calculated the MAE between the cGAN-models output, and the original image corresponding to the label used as input. In the second method, we used a pre-trained convolutional regression model to predict the spectral data of a nanostructure image. This approach allows us to estimate the label error instead of the pixel error, by calculating the MAE on the output from the regression model. To estimate the mean value of the given metric, we drew $n$=30 stochastic samples from the cGAN-model and considered the average.

Figure \ref{fig:2} presents an example prediction from the FCGAN-model on test data not seen at all during the model development. The FEM simulation data in \cref{fig:2}d and \cref{fig:2}e are the conditional inputs used to generate the output displayed in \cref{fig:2}b and \cref{fig:2}c. We used the pre-trained regression model to evaluate the model's performance, by predicting corresponding spectral data of the output images. The evaluated spectral data are plotted together with the FEM-simulated data in \cref{fig:2}d and \cref{fig:2}e. Adding label projection and the embedding network to the model improves image quality, and consequently, the estimated spectral data lie closer to the FEM-simulated data. This is especially true for the scattering cross section in this specific example.

From the results in \cref{fig:3}, we find for both network architectures that the convergence time for training the model decreases when label projection is added. We believe the label projection helps the optimization of the model parameters as shown by \cite{miyato2018cgans}, leading to fewer epochs required for convergence. Adding the embedding network, also speeds up the convergence, but the effect is smaller compared to adding label projection. This results in faster training of the models which is important for such computationally heavy algorithms. It should be noted that the embedding network increases the training time per epoch by approximately 17\%, compared to only using the label projection. This has to be considered when selecting the best version of the model as the embedding network slows down training, but in some cases improves error estimates.

We also find that the performance, in terms of error estimates, improves for the FCGAN-model. The improvement is present, both for adding label projection and the embedding network, but the combination of the two results in the smallest error. This suggests that the model needs more capacity since it benefits from the additional parameters added in the embedding network. For the DCGAN-model, the model performance reaches what looks as a lower limit for the model on this dataset. The error converges to roughly the same value for the 4 variants, which indicates that there is enough capacity in the models for learning from this dataset. However, the DCGAN-model + LP + Embed. converges in about 5000 epochs, which is more than 3 times faster compared to the standard DCGAN-model. All models in the study were trained for 30000 epochs, and this leads to over-fitting for the variants that converge faster. This is expected as they reach their optimum earlier, which results in overtraining in the remaining epochs. 

\begin{figure*}[htb!]
    \centering
    \includegraphics[width=0.65\linewidth]{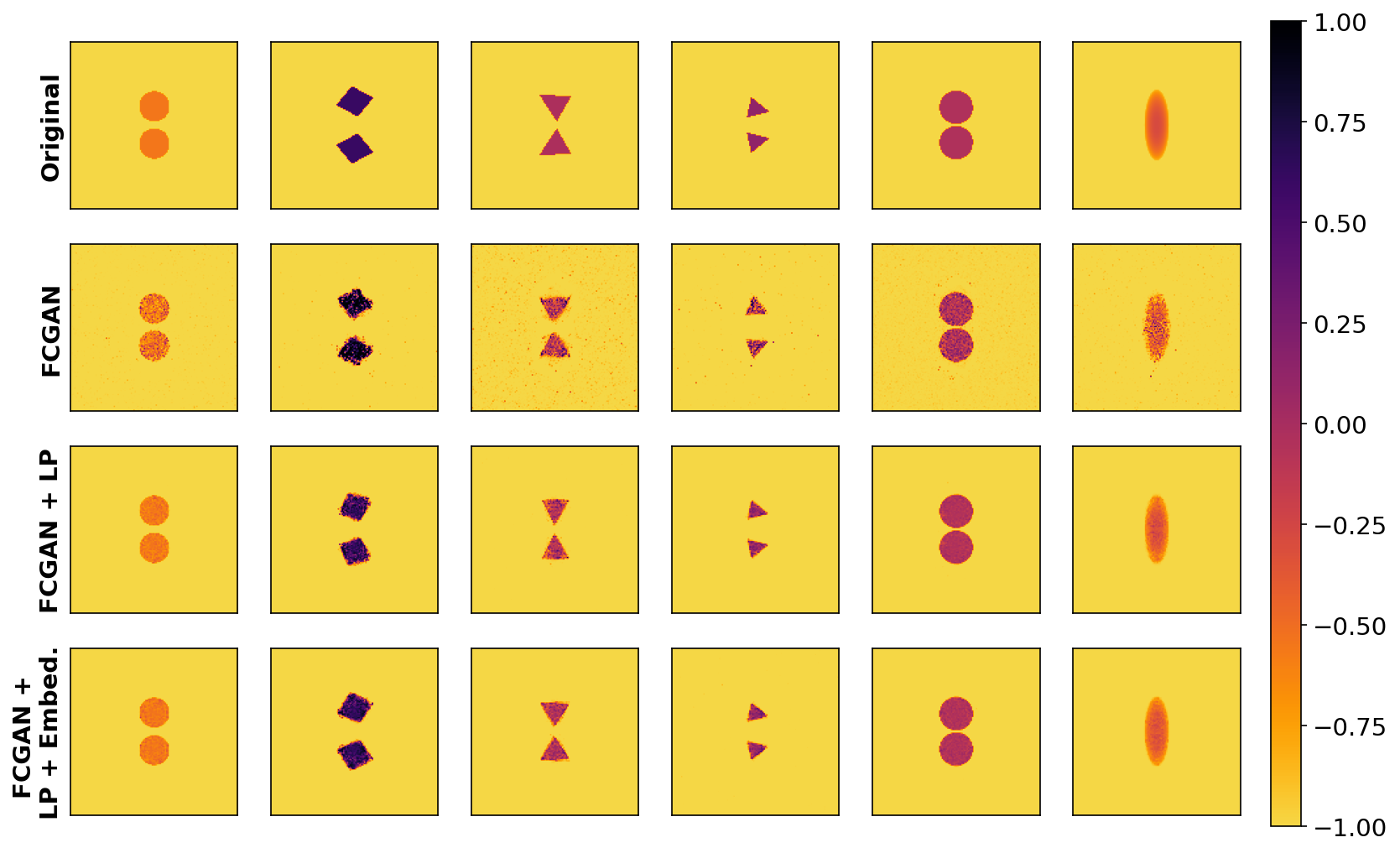}
    \caption{This figure shows example images from training the FCGAN-models on a larger dataset containing differently shaped anisotropic gold structures. The first row contains the original images, and the other rows contains the corresponding predicted images by the FCGAN-models. The model predictions improve when label projection and the embedding network are added to the model.}
    \label{fig:4}
\end{figure*}

In addition, the FCGAN-models were trained on another dataset, containing structures with a more complex shape. Figure \ref{fig:4} illustrates predictions on test data for three versions of the FCGAN-model. These results demonstrate that our models can also learn from datasets with more complex shapes as well, and all the predictions improved by adding the LP + Embed. into the model. 

It is clear from the training results on the cylindrical dimer dataset in \cref{fig:3}, that both the FCGAN and DCGAN model benefit from our proposed improvements. The DCGAN-model sees a significant decrease in convergence time as label projection is added to the critic architecture, and adding the embedding network improves even more the performance. However, the models produce similar error estimates across the four versions. This indicates that the deep convolutional architecture in itself has a sufficient number of trainable parameters to capture the feature of the dataset, leading to a near-optimal solution in each case. 

Notably, for the FCGAN-model, we found that the combination of using the label embedding network with the label projection gives the best performance, in terms of the number of epochs required for convergence and error estimates. These results are well illustrated in \cref{fig:4} and \cref{fig:2} where predictions from the FGCAN-model are shown. The image quality improves significantly as the label projection is added, leading to a better estimate of the optimal nanostructure design. This is an important finding, as it shows that a simple model without convolutional layers can be improved with small modifications to a low cost in computational resources. Compared to the more computationally expensive DCGAN model, its performance is not that far off in terms of error estimates, particularly when considering corresponding cross-section spectra in \cref{fig:3}c and \cref{fig:3}d. In practice, training would have been stopped before the models starts to overfit, ensuring the lowest possible error estimate is achieved. Therefore, a fair comparison should evaluate the two models at their respective optimal error estimates, that is the epoch where the lowest error is achieved. In doing so, we conclude that the best FCGAN model is performing well in terms of the MAE on cross-section spectra compared to the best DCGAN model. However, the DCGAN models are slightly better when it comes to the MAE of the actual images, since the convolutional layers in the DCGAN model helps to generate higher quality images.

\subsection*{Challenges and Limitations}
\begin{figure*}[htb!]
    \centering
    \includegraphics[width=0.75\linewidth]{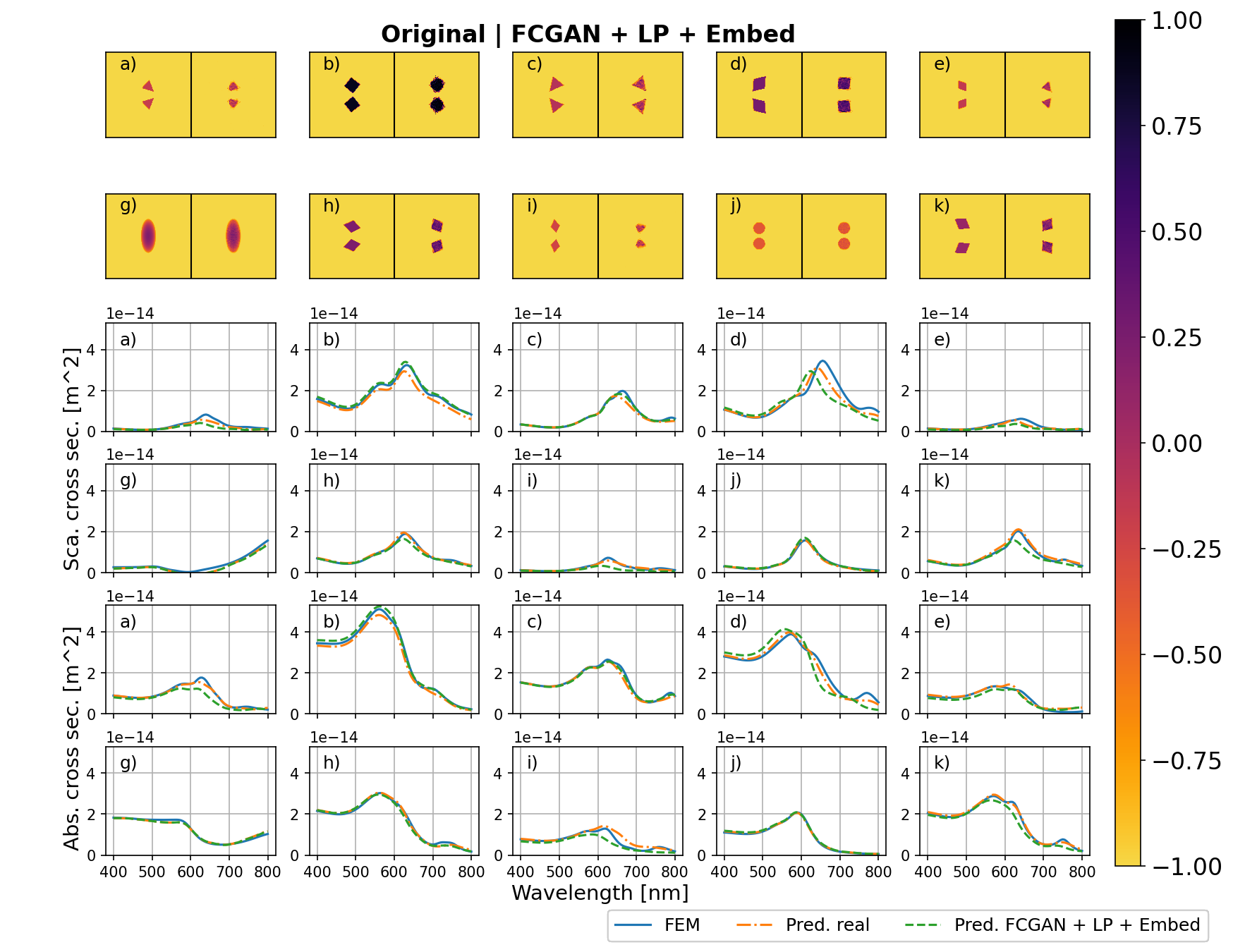}
    \caption{The figure shows images generated from the FCGAN + LP + Embedding network model and the corresponding spectral data predicted by a pretrained regression model, just as in \cref{fig:2}. Original images are to the left, and generated ones are to the right. The samples are chosen to illustrate the fact that one specific cross section sprectra might correspond to several nanostructures designs, the so called one-to-many problem. As a result, the GAN-model output can differ from the original image while its corresponding spectra are still closely related. For instance, the model outputs images that are mirrored compared to the original image. This is likely explained by the symmetry of the structures with respect to the polarization vector, that makes a 45 degree angle with the long axis of the dimers. Images c), and k) illustrate the phenomenon. We also observe cases where the model outputs structures with where the shape does not match exactly. Often the cross section spectra is still similar, as seen from a), b), d), e), h), i), j).}
    \label{fig:5}
\end{figure*}

\noindent With a growing design parameter space, there may exist more than one solution to the conditional input provided. In general, one cross-section spectra may correspond to several structures in the design space, and this fact is illustrated in \cref{fig:5}. The examples are chosen to illustrate the non-uniqueness of solutions to the inverse design problem, which makes the problem particularly challenging when designing an inverse design model. Several solutions may exist for a single input, possibly leading to parts of the solution space not being reachable by the model \cite{dai2022inverse}. Conditional GAN models are well suited to tackle this problem as the stochastic input allows multiple solutions to be learned, and as seen in \cref{fig:5}, the FCGAN-model with label projection and embedding network, successfully learns to output structures with a similar cross-sectional spectra as the original one. In many cases the output is a 180 degree rotation of the original structure, due to the symmetry of the problem with respect to the polarisation vector of the linearly polarised light illuminating the structures. This is fine since parity symmetry is not violated, but still gives an image different from the original one. Other predictions include structures where the output shape does not match exactly the original one, but the cross-sections spectra are still similar which is what we want to achieve. The model fails to make accurate predictions for some inputs, typically the smaller structures where the image resolution potentially starts to become a limiting factor. Even in this case the main optical features are still captured. It seems that the size of the predicted structures as well as the gap between the dimers, are consistently well estimated, and this leads to reasonably small pertubations in the cross-sections spectra for the examples in \cref{fig:5}.

The use of the Wasserstein GAN training algorithm and loss function should also be discussed as it has its own strengths and weaknesses. Most importantly, it has a stabilizing effect on the optimization process since the loss function provides smoother gradients, which avoids instability problems such as vanishing and exploding gradients. Those problems are preferably avoided when optimizing GAN-models, since the convergence of the algorithm become very sensitive to the hyperparameter values, particularly the learning rate. Using the Wasserstein loss comes at the cost though, because the computational load increases, as the penalty term in \cref{eq:wgan} requires an additional gradient computation in each iteration, which leads to slightly longer training time.
\newpage

\subsection*{Conclusion}
\noindent Conditional generative adversarial networks have recently emerged as popular inverse design models within the nanophotonic community. In this study, we have shown how two standard models can be modified to improve the convergence of the training algorithm and the accuracy of the predicted results. Most importantly, these results are achieved through pure algorithmic improvements without adding a large amount of new parameters to the model architecture. This keeps the cost of optimizing the models down, both in terms of training and memory usage. We believe this is an important step towards more efficient inverse design models for nanophotonic applications.

\subsection*{Code and Data availability}
\noindent The code and all data that support the findings in this study are publicly available in the git repository \url{https://github.com/pettper/Improving-conditional-generative-adversarial-networks-for-inverse-design-of-plasmonic-structures}

%% file: chapters/acknowledgments.tex
\noindent This work had been funded by Swedish Research Council (Grant No. 2021-05784), the Knut and Alice Wallenberg Foundation through the Wallenberg Academy Fellows Program (Grant No. 2023.0089) and the European Research Council through the ERC Starting Grant ‘MagneticTWIST’ (Grant No. 101116253).

We thank the High Performance Computing Center North (HPC2N) at Umeå University for providing computational resources and valuable support during test and performance runs.

%% file: chapters/supplementary.tex
\renewcommand{\thetable}{S.\arabic{table}}  
\setcounter{table}{0}
\renewcommand{\thefigure}{S.\arabic{figure}}  
\setcounter{figure}{0}
\label{sec:supp}
\noindent Figures \ref{fig:sup_upsampling} and \ref{fig:sup_downsampling} show schematics of the neural networks used in the proposed inverse design model for downsampling and upsampling of images. In the DCGAN-model we use the convolutional architecture proposed by \myCite{radford2015unsupervised} modified to suite the label projection used in our models. This architecture utilizes the two-dimensional convolutional operation for downsampling and the corresponding transposed convolutional operation for upsampling. In the DCGAN downsampling block, we use instance normalization \cite{ulyanov2016instance} instead of batch normalization, to avoid introducing correlations between samples. This makes the gradient penalty work properly as suggested by as suggested by \cite{gulrajani2017improved}. The LeakyReLU activation function with a slope of 0.2 is applied after all hidden layers in the downsampling block. In the DCGAN upsampling block we use batch normalization \cite{ioffe2015batch} together with ReLU as activation function after the transposed convolutions in the hidden layers. In the output layer, the tangent hyperbolic function is used as activation function as in \cite{radford2015unsupervised} which requires the images to be normalized to the range [-1, 1]. The details of the layers are showed in \cref{tab:sup_dcdisc} and \cref{tab:sup_dcgen} for 128x128 images.
\begin{figure}[htb!]
    \centering
    \includegraphics[width=0.8\linewidth]{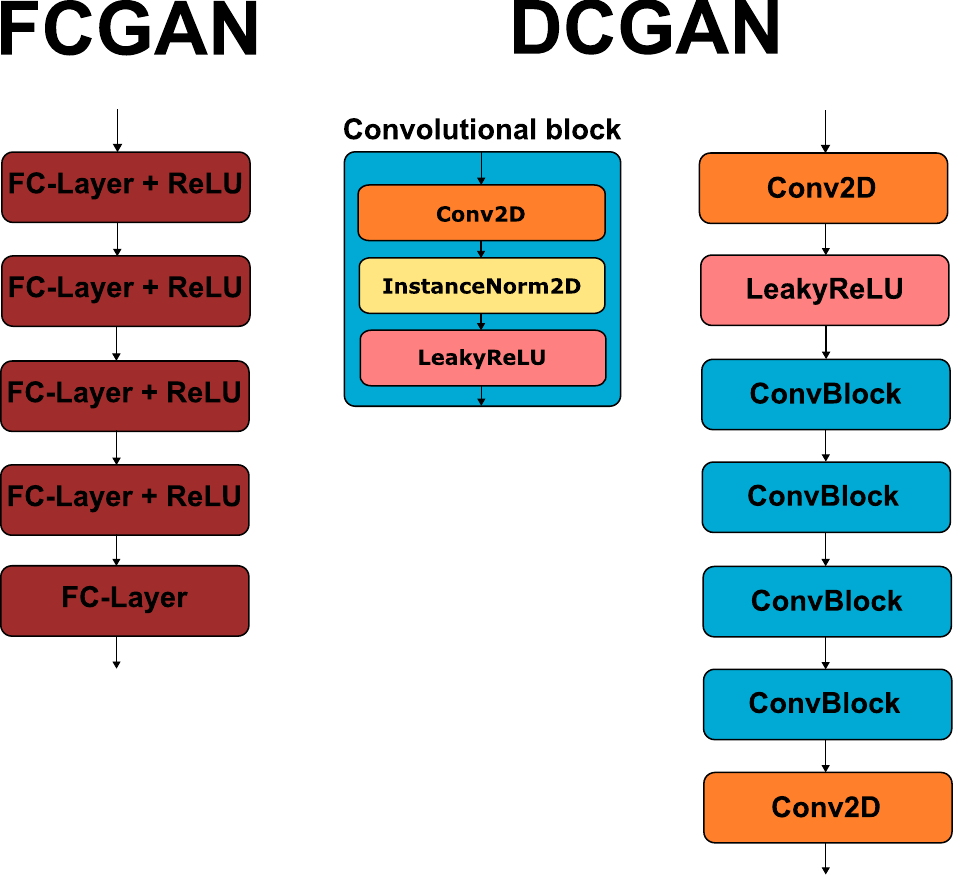}
    \caption{The figure shows the FCGAN and DCGAN downsampling blocks.}
    \label{fig:sup_downsampling}
\end{figure}
\begin{figure}[htb!]
    \centering
    \includegraphics[width=0.8\linewidth]{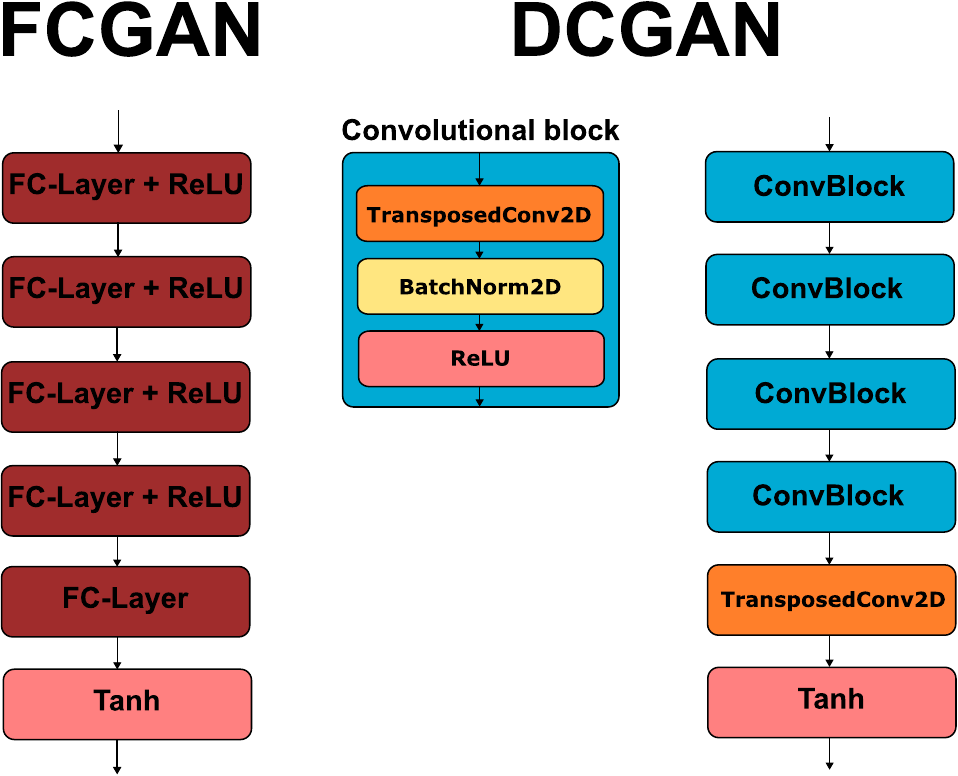}
    \caption{The figure shows the FCGAN and DCGAN architectures for the image upsampling blocks.}
    \label{fig:sup_upsampling}
\end{figure}
\begin{table}[htb!]
\centering
\caption{The table summarize the critic architecture for the DCGAN model.}
\resizebox{\linewidth}{!}{%
\begin{tabular}{|c|c|c|c|c|}
\hline
Stage & Operation & \begin{tabular}[c]{@{}c@{}}Channels\\ in $\rightarrow$ out\end{tabular} & \begin{tabular}[c]{@{}c@{}}Kernel\\ /Stride\\ /Padding\end{tabular} & \begin{tabular}[c]{@{}c@{}}Image Size\\ in $\rightarrow$ out\end{tabular} \\ \hline
1 & \begin{tabular}[c]{@{}l@{}}Conv\\ + LeakyReLU(0.2)\end{tabular} & 2 $\rightarrow$ 64 & 4/2/1 & 128×128 $\rightarrow$ 64×64 \\ \hline
2 & \begin{tabular}[c]{@{}l@{}}Conv\\ + InstanceNorm\\ + LeakyReLU(0.2)\end{tabular} & 64 $\rightarrow$ 128 & 4/2/1 & 64×64 $\rightarrow$ 32×32 \\ \hline
3 & \begin{tabular}[c]{@{}l@{}}Conv\\ + InstanceNorm\\ + LeakyReLU(0.2)\end{tabular} & 128 $\rightarrow$ 256 & 4/2/1 & 32×32 $\rightarrow$ 16×16 \\ \hline
4 & \begin{tabular}[c]{@{}l@{}}Conv\\ + InstanceNorm\\ + LeakyReLU(0.2)\end{tabular} & 256 $\rightarrow$ 512 & 4/2/1 & 16×16 $\rightarrow$ 8×8 \\ \hline
5 & \begin{tabular}[c]{@{}l@{}}Conv\\ + InstanceNorm\\ + LeakyReLU(0.2)\end{tabular} & 512 $\rightarrow$ 1024 & 4/2/1 & 8×8 $\rightarrow$ 4×4 \\ \hline
6 & Conv & 1024 $\rightarrow$ 50 & 4/2/0 & 4×4 $\rightarrow$ 1×1 \\ \hline
\end{tabular}%
}
\label{tab:sup_dcdisc}
\end{table}

\begin{figure*}[t]
    \centering
    \includegraphics[width=0.7\linewidth]{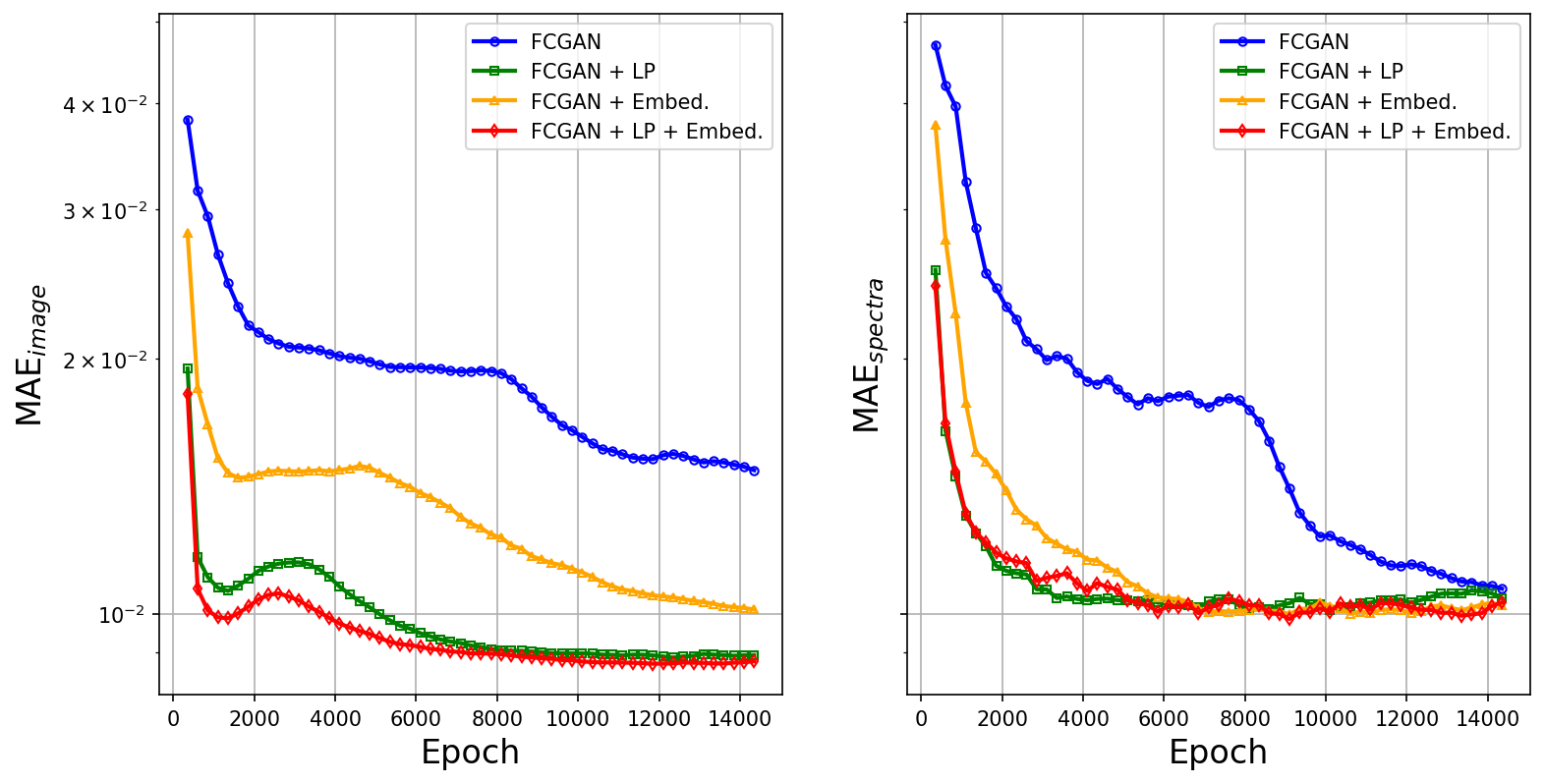}
    \caption{The models were trained on a dataset containing anisotropic gold nano structures and they were evaluated using two different methods. The first method uses a pre-trained convolutional neural network regression model to predict the scattering cross section and absorption cross section spectra. This approach allows us to estimate the mean absolute error between the spectra predicted from real and GAN-generated images. As a second evaluation method, we also estimate the mean absolute error in pixel values between real and generated images. Figures a) and b) shows the results from training the FCGAN architectures. We calculate the evaluation metrics by generating n=30 sample images for each labeled original image, and take the average of those when computing the error estimate.}
    \label{fig:sup_error_plots_anisotropic}
\end{figure*}

\begin{table}[htb!]
\centering
\caption{The table summarize the generator architecture for the DCGAN model.}
\resizebox{\linewidth}{!}{%
\begin{tabular}{|c|c|c|c|c|}
\hline
Stage & Operation & \begin{tabular}[c]{@{}c@{}}Channels\\ in $\rightarrow$ out\end{tabular} & \begin{tabular}[c]{@{}c@{}}Kernel\\ /Stride\\ /Padding\end{tabular} & \begin{tabular}[c]{@{}c@{}}Image Size\\ in $\rightarrow$ out\end{tabular} \\
\hline
1 & \begin{tabular}[c]{@{}c@{}}ConvTranspose\\ + BatchNorm\\ + ReLU\end{tabular} & 1024 $\rightarrow$ 512 & 4/2/1 & 4×4 $\rightarrow$ 8×8 \\
\hline
2 & \begin{tabular}[c]{@{}c@{}}ConvTranspose\\ + BatchNorm\\ + ReLU\end{tabular} & 512 $\rightarrow$ 256 & 4/2/1 & 8×8 $\rightarrow$ 16×16 \\
\hline
3 & \begin{tabular}[c]{@{}c@{}}ConvTranspose\\ + BatchNorm\\ + ReLU\end{tabular} & 256 $\rightarrow$ 128 & 4/2/1 & 16×16 $\rightarrow$ 32×32 \\
\hline
4 & \begin{tabular}[c]{@{}c@{}}ConvTranspose\\ + BatchNorm\\ + ReLU\end{tabular} & 128 $\rightarrow$ 64 & 4/2/1 & 32×32 $\rightarrow$ 64×64 \\
\hline
5 & ConvTranspose & 64 $\rightarrow$ 2 & 4/2/1 & 64×64 $\rightarrow$ 128×128 \\
\hline
6 & Tanh & 2 $\rightarrow$ 2 & – & 128×128 \\
\hline
\end{tabular}%
}
\label{tab:sup_dcgen}
\end{table}
In the FCGAN-model we use fully connected layers with a ReLU activation function between each layer. The hidden layers contain 512 neurons each. The upsampling and downsampling blocks are similar in their internal and differ only in the input and output layers. The tangent hyperbolic function is used as output activation in the upsampling block, just as in the DCGAN-model. The details of the layers are showed in \cref{tab:sup_fcdisc} and \cref{tab:sup_fcgen} for 128x128 images. Figure \ref{fig:sup_error_plots_anisotropic} shows the validation error from our metrics after training the four different versions of the FCGAN-model. We use a dataset containing anisotropic gold nano structures and train the models for 15000 epochs. 

\begin{figure*}[t]
    \centering
    \includegraphics[width=0.6\linewidth]{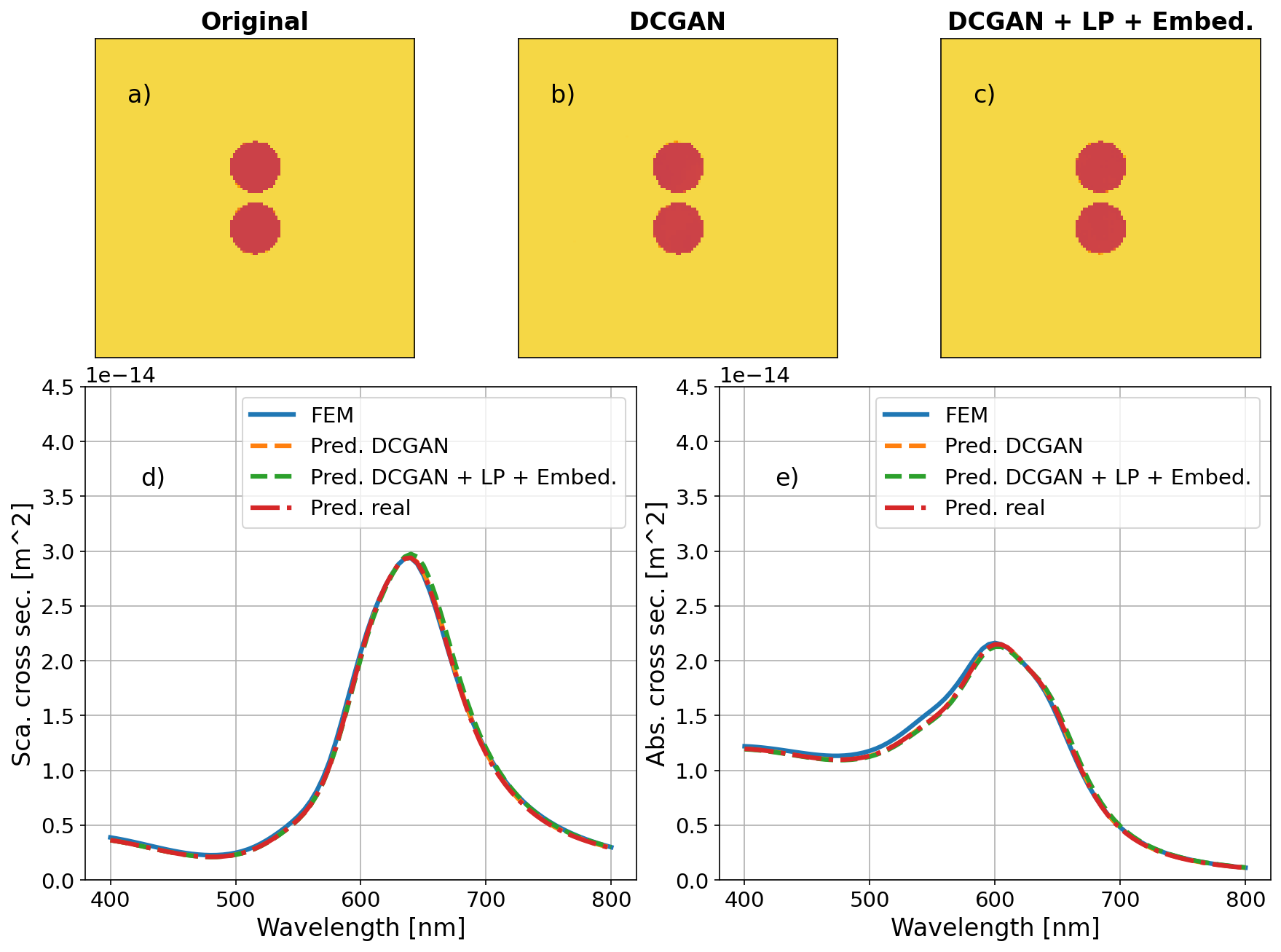}
    \caption{The results in this figure are obtained from the DCGAN-model with and without using label projection and an embedding network in the critic. Figure a) show the original image and figures b) and c) shows the images predicted by the models. Figures d) and e) show the scattering and absorption cross section spectras corresponding to the images. The blue curve is the original data from the finite element simulation of the gold nano structure in the original image. We use a pretrained convolutional neural network regression model to predict the spectral data for both the original image (dashed red) and the images generated by the model (dashed orange and green). The results are similar in terms of image quality and estimated cross section spectrum for both variants of the model, and the predicted designs agree well with the original structure.}
    \label{fig:sup_dcgan_example}
\end{figure*}

The DCGAN-model was also trained on the dataset with anisotropic gold nano structures and \cref{fig:sup_dcgan_example} shows one example prediction after training for 30000 epochs. We see that there is excellent agreement with the original image both with and without label projection. The estimation of the spectral data, also show very good agreement with the original simulation data.
\begin{table}[htb!]
\centering
\caption{The table summarize the critic network for the FCGAN-model.}
\newcolumntype{C}{>{\centering\arraybackslash}X}
\begin{tabularx}{\linewidth}{|c|C|c|c|}
\hline
Stage & Operation & Input dim & Output dim \\ \hline
1 & FC-Layer + ReLU & 32768 & 512 \\ \hline
2 & FC-Layer + ReLU & 512 & 512 \\ \hline
3 & FC-Layer + ReLU & 512 & 512 \\ \hline
4 & FC-Layer + ReLU & 512 & 512 \\ \hline
5 & FC-Layer & 512 & $d_e$ \\ \hline
\end{tabularx}
\label{tab:sup_fcdisc}
\end{table}
\begin{table}[htb!]
\centering
\caption{The table summarize the generator network for the FCGAN-model.}
\newcolumntype{C}{>{\centering\arraybackslash}X}
\begin{tabularx}{\linewidth}{|c|C|c|c|}
\hline
Stage & Operation & Input dim & Output dim \\ \hline
1 & FC-Layer + ReLU & $d_e$ & 512 \\ \hline
2 & FC-Layer + ReLU & 512 & 512 \\ \hline
3 & FC-Layer + ReLU & 512 & 512 \\ \hline
4 & FC-Layer + ReLU & 512 & 512 \\ \hline
5 & FC-Layer & 512 & 32768 \\ \hline
6 & Tanh & 32768 & 32768 \\ \hline
\end{tabularx}
\label{tab:sup_fcgen}
\end{table}


\renewcommand{\thetable}{\arabic{table}}